\documentclass[10pt,twocolumn]{IEEEtran}
\usepackage{graphicx}           
\usepackage{amsmath,amsfonts}
\usepackage{times,epsfig}
\usepackage{psfrag}
\usepackage{stfloats}
\usepackage{amssymb}
\usepackage{multirow}
\usepackage{cite}
\usepackage{CJK}
\usepackage{url}
\usepackage{caption}
\usepackage{array}
\usepackage{soul}

\usepackage{booktabs,ragged2e}
\usepackage[flushleft]{threeparttable}
\usepackage{tabularx}
\usepackage{makecell}
\usepackage{bm}
\usepackage{color}
\usepackage{hyperref}

\begin{document}
\title{Toward Resource-Efficient Collaboration of Large AI Models in Mobile Edge Networks}

\author{
Peichun~Li, Liping Qian,~\IEEEmembership{Senior~Member,~IEEE}, Dusit Niyato,~\IEEEmembership{Fellow,~IEEE}, \\Shiwen Mao,~\IEEEmembership{Fellow,~IEEE}, and Yuan Wu,~\IEEEmembership{Senior~Member,~IEEE}\vspace{-5pt}

\IEEEcompsocitemizethanks{
\IEEEcompsocthanksitem P. Li and Y. Wu are with the State Key Laboratory of Internet of Things for Smart City and the Department of Computer Information Science, University of Macau, Macao SAR, China (e-mail: peichun.li@connect.um.edu.mo, yuanwu@um.edu.mo).
\IEEEcompsocthanksitem L. Qian is with Institute of Cyberspace Security, Zhejiang University of Technology, Hangzhou 310023, China (e-mail: lpqian@zjut.edu.cn).
\IEEEcompsocthanksitem D. Niyato is with the College of Computing \& Data Science (CCDS), Nanyang Technological University, Singapore, Block N4-02a-32, Nanyang Avenue, Singapore. (e-mail: dniyato@ntu.edu.sg).
\IEEEcompsocthanksitem S. Mao is with the Department of Electrical and Computer Engineering, Auburn University, Auburn, AL 36849-5201, USA. (e-mail: smao@ieee.org).
\IEEEcompsocthanksitem \textit{Yuan Wu is the corresponding author.}}
}

\maketitle \thispagestyle{headings}

\begin{abstract}
The collaboration of large artificial intelligence (AI) models in mobile edge networks has emerged as a promising paradigm to meet the growing demand for intelligent services at the network edge. By enabling multiple devices to cooperatively execute submodels or subtasks, collaborative AI enhances inference efficiency and service quality with constrained resources. However, deploying large AI models in such environments remains challenging due to the intrinsic mismatch between model complexity and the limited computation, memory, and communication resources in edge networks. This article provides a comprehensive overview of the system architecture for collaborative AI in mobile edge networks, along with representative application scenarios in transportation and healthcare. We further present recent advances in resource-efficient collaboration techniques, categorized into spatial and temporal approaches. The major spatial approaches include federated tuning, mixture of experts, patch-based diffusion, and hierarchical diffusion. Meanwhile, the important temporal approaches encompass split learning, cascading inference, speculative decoding, and routing inference. Building upon these foundations, we propose a multi-stage diffusion framework that enables elastic distribution of large generative models across heterogeneous edge resources. Experimental results demonstrate that our framework achieves performance improvement in both efficiency and adaptability for data generation.
\end{abstract}
           

\section{Introduction}
Large artificial intelligence (AI) models, including large language models (LLMs), vision language models (VLMs), and generative diffusion models (GDMs),  have demonstrated remarkable capabilities across a wide range of applications. By leveraging these models, mobile edge networks can provide intelligent and adaptive services, such as intelligent transportation and personalized healthcare \cite{10591707}. However, the complexity and computational intensity of large AI models pose severe challenges for traditional cloud-based deployments. Centralized execution often leads to high latency and excessive backbone traffic, thereby undermining the responsiveness and reliability demanded by emerging edge applications.

Mobile edge networks have been evolving to accommodate the increasing intelligence requirements at the network edge. Shifting computation closer to end users reduces reliance on remote cloud resources and enables low-latency interactions with mobile applications. Unlike centralized cloud infrastructures, edge networks are inherently heterogeneous, consisting of mobile devices and edge servers connected via wireless links. This heterogeneity offers both opportunities and challenges: it allows exploitation of underutilized local computation and communication resources, while simultaneously complicating the coordination of heterogeneous capabilities across distributed devices.

Recently, the collaboration of large AI models has emerged as a promising paradigm \cite{10759588}. Instead of relying on a centralized server, collaborative AI distributes submodels and subtasks across client devices, edge servers, and cloud infrastructures. Coordinated allocation of computation, communication, and memory resources enables distributed entities to collaboratively deliver high-quality intelligent services. This paradigm not only improves efficiency by exploiting parallelism but also increases robustness through redundancy and adaptability. Consequently, collaborative AI provides a foundation for deploying large models in mobile edge networks.

However, the integration of large AI models with mobile edge networks remains challenging. On the one hand, the structural complexity of modern models hinders the effective decomposition of the whole model into modules, making it difficult to achieve distributed execution while preserving accuracy \cite{patel2024splitwise}. On the other hand, limited computation and communication resources at edge devices restrict the edge devices to handle intensive workloads \cite{xu2024dynamic}. This tradeoff necessitates a joint optimization between the large AI model and edge networks. Such an optimization requires cross-layer strategies that simultaneously consider AI model design and system resources to ensure efficient and reliable deployment.

Existing studies have explored this problem from different perspectives. The communication and networking community has primarily emphasized resource optimization, including task scheduling and bandwidth allocation \cite{10591707, kang2025diffusion}, yet large AI models are often treated as closed boxes. In contrast, the AI community has advanced techniques such as model compression and efficient inference, but often overlooks the constraints of wireless communication and heterogeneous edge infrastructures. This disciplinary gap hinders the full realization of collaborative AI in mobile edge networks.

In this article, we investigate the integration of large AI models with mobile edge networks, with a focus on enhancing the efficiency of intelligent services through multi-device collaboration. To this end, we first outline the four-layer system architecture of collaborative AI systems. We then review existing collaborative AI approaches to highlight their benefits and limitations. After that, we propose a case study with multi-stage diffusion to improve the efficiency of the data generation process. We summarize the detailed contributions in this article as follows.
\begin{itemize}
    \item We propose a comprehensive system architecture that facilitates the deployment of large AI models in mobile edge networks. This architecture outlines the interactions between applications, AI models, optimization techniques, and the edge resources, offering a structured framework for integrating diverse edge resources.
    \item We conduct an in-depth review of existing resource-efficient collaboration strategies, categorizing them into spatial and temporal approaches. We analyze the potential of these strategies to optimize computational and communication resources, highlighting their benefits and limitations in collaborative networks.
    \item We introduce a multi-stage relay diffusion framework for diffusion models, which decomposes the diffusion process into multiple stages and allocates tasks across heterogeneous edge devices. Additionally, we propose a device selection scheme that can filter out weak devices for accelerating the overall generation process.
\end{itemize}

\begin{figure*}[t]\centering
  \includegraphics[width=0.86\textwidth]{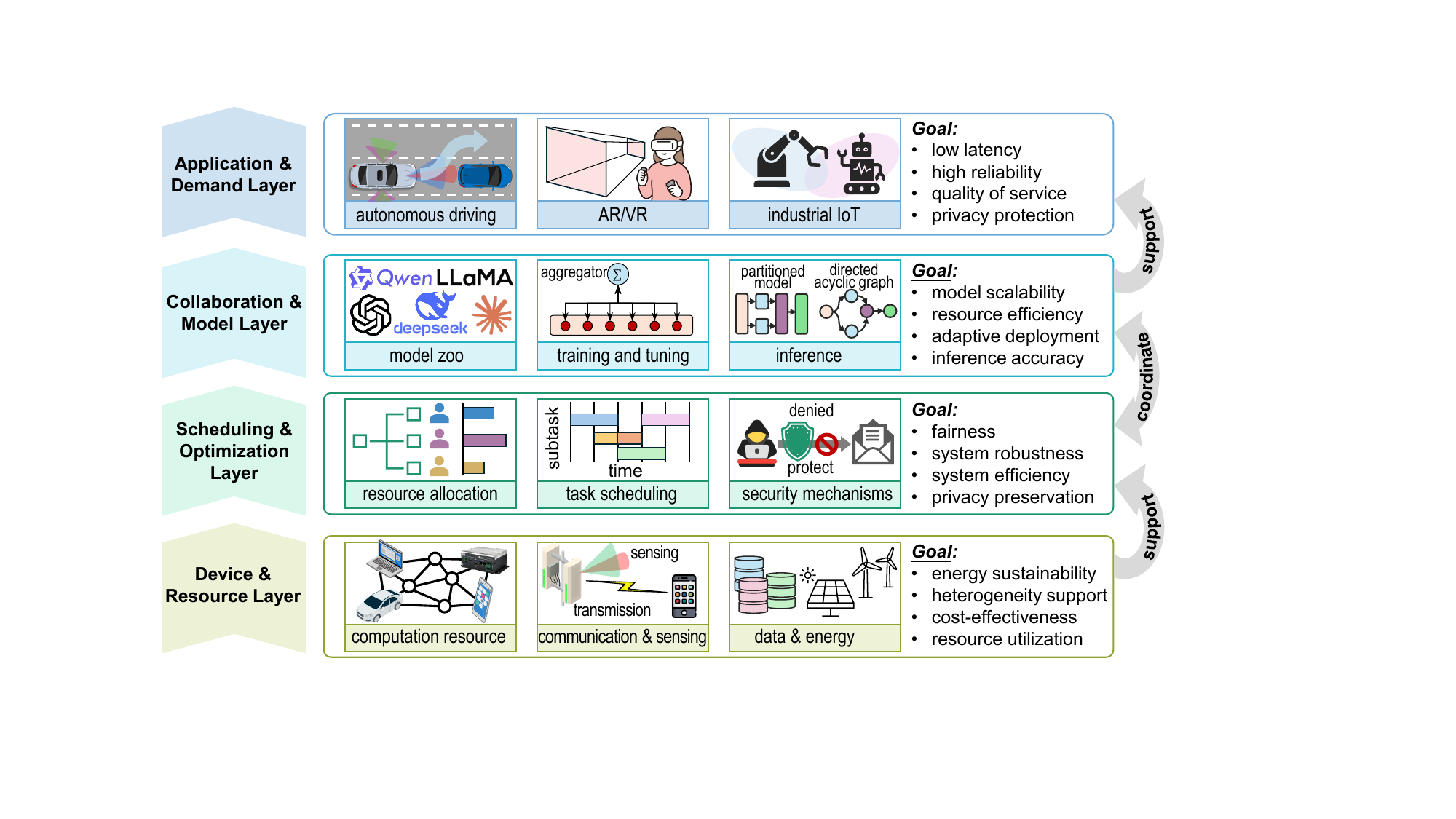}
  \caption{System architecture of collaborative AI in mobile edge networks. The system consists of four layers: the application and demand layer defines service requirements; the collaboration and model layer enables scalable and adaptive model execution; the scheduling and optimization layer coordinates AI tasks and resources; and the device and resource layer provides the underlying infrastructure. Higher layers specify demands, while lower layers offer resource and algorithm support.}\label{archit}
  \vspace{-6pt}
\end{figure*}

The remainder of this article is organized as follows. Section II introduces the overall framework of collaborative AI in mobile edge networks. Section III reviews the existing resource-efficient collaboration approaches. Section IV presents the multi-stage relay diffusion framework. Section V discusses key open challenges, and Section VI finally concludes the article.

\section{Overview of Collaborative AI in Edge Networks}
\subsection{System Architecture}
As illustrated in Figure~\ref{archit}, the system architecture of collaborative AI in edge networks is organized into four layers. The functionalities of these layers are summarized as follows.
\subsubsection{Application and Demand Layer}
The application and demand layer defines the service objectives that drive the collaborative deployment of large AI models. Different applications impose varying requirements on key factors such as latency, accuracy, privacy, and reliability. For instance, autonomous driving systems prioritize ultra-low latency and high reliability, while immersive media applications focus on high-quality and low-latency streaming. These diverse demands translate into critical system constraints, which in turn shape the design of collaborative AI solutions.

\subsubsection{Collaboration and Model Layer}
The collaboration and model layer facilitates distributed execution across mobile edge networks, which is crucial for deploying large AI models. This layer leverages advanced neuron-aware techniques to achieve resource-efficient training and inference that balance high performance with low latency. By structuring AI models into modular and compressible components, this layer enables the dynamic distribution of computational tasks across various devices, which further establishes the foundation for optimizing resource allocation in lower layers.

\subsubsection{Scheduling and Optimization Layer}
The scheduling and optimization layer provides the orchestration mechanisms needed to translate collaborative AI designs into real-world deployments. It includes strategies for scheduling tasks, optimizing bandwidth allocation, implementing communication-efficient learning algorithms, and ensuring privacy protection. Through dynamic coordination of computation and communication resources across diverse participants \cite{10454003}, this layer ensures that collaborative AI systems can adapt to network variability and device heterogeneity \cite{xu2024enhancing}.

\subsubsection{Device and Resource Layer}
At the base of the architecture, the device and resource layer represents the physical infrastructure of mobile devices, edge servers, and communication links. This layer provides the computational, storage, and energy resources essential for supporting collaborative AI models. The diversity of devices enables distributed models such as multi-edge collaboration and displaced data and model parallelism, grounding collaborative AI in the practical realities of mobile edge networks.

\subsection{Representative Application Scenarios}
\begin{figure*}[t]\centering
  \includegraphics[width=0.88\textwidth]{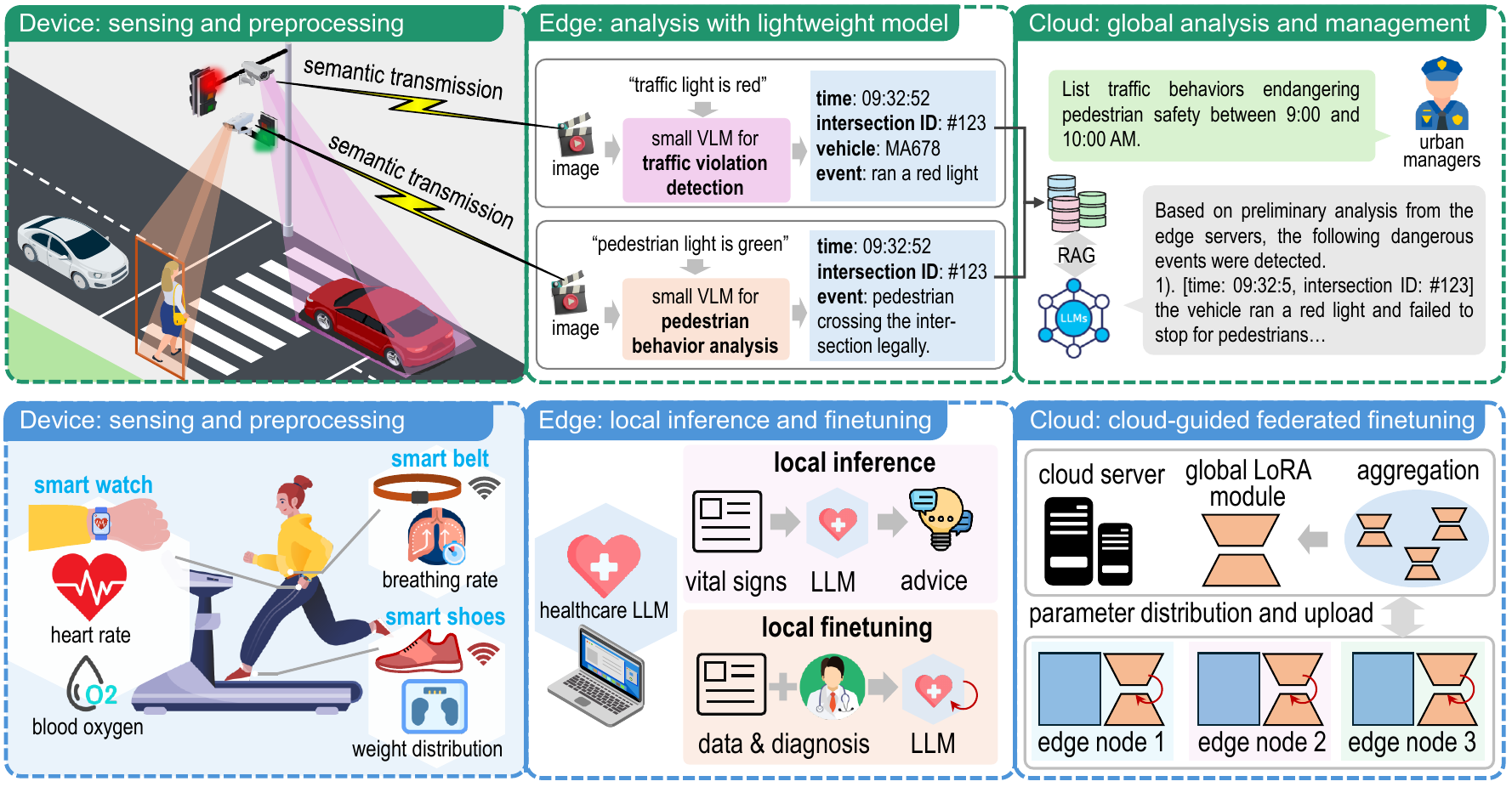}
  \caption{Typical application scenarios of collaborative AI in mobile edge networks. \textbf{Top}: In transportation, roadside sensors collect traffic data, edge servers with lightweight VLMs identify traffic behaviors, and the cloud-based LLM aggregates results for global safety management. \textbf{Bottom}: In healthcare, wearable sensors monitor vital signs, edge devices enable local LLM inference and lightweight finetuning, and the cloud serves as the parameter server for federated finetuning.}\label{appli}
  \vspace{-8pt}
\end{figure*}

We provide representative application scenarios of collaborative AI in mobile edge networks, as shown in Figure~\ref{appli}. {These scenarios demonstrate how distributed models and resources can be jointly leveraged to support intelligent services in domains such as transportation\footnote{Vialytics. \url{https://www.vialytics.com/}} and healthcare\footnote{Feel Therapeutics. \url{https://www.feeltherapeutics.com/}}.}

\subsubsection{Intelligent Road Monitoring} The intelligent road monitoring is an essential example of edge-enabled sensor networks in transportation. It involves multi-layer collaboration between client devices, edge servers, and cloud infrastructure to ensure efficient traffic management.
\begin{itemize}
    \item Roadway sensors are deployed along roads, highways, and intersections to continuously monitor traffic conditions. These client devices can utilize semantic communication to upload the data to edge processing units.
    \item Multiple edge processing units are deployed with specialized small VLMs for different tasks, including license plate recognition, traffic violation detection, and pedestrian behavior analysis. Each edge unit acts as an expert for local inference on a specific task and sends the results to the cloud server.
    \item A cloud server plays as the centralized platform for decision-making with LLM. Urban managers can query the cloud server using text prompts, and the cloud can employ multi-modal retrieval-augmented generation (RAG) techniques to fetch relevant video footage.
\end{itemize}

\subsubsection{Wearable Sensors for Healthcare}
In healthcare, wearable sensors play a critical role in continuously monitoring a user's vital signs, providing personalized healthcare guidance. This involves local inference and federated training for specialized large models.
\begin{itemize}
    \item Wearable devices with heart rate monitors and glucose sensors collect real-time data on a user’s vital signs. These devices perform basic local processing to detect immediate irregularities while transmitting the sensing data to dedicated edge devices via Wi-Fi and Bluetooth.
    \item The edge device, typically a personal computer or dedicated local gateway, handles two key tasks:
    i) It hosts a local large model for more complex inferences, producing short-term health suggestions based on the user's daily vital signs.
    ii) It acts as a training node, adapting and tuning the local model instructed by the cloud.
    \item The cloud server guides the edge devices in federated learning, enabling multiple nodes to refine the domain-specified LLM. Efficient techniques such as low-rank adaptation can be employed to reduce the training overhead.
\end{itemize}

\subsection{Key Challenges}
\subsubsection{Challenges from the Large AI Model}
To enable collaborative execution, large models are often partitioned into smaller components that can be distributed across multiple devices. Partitioning can be organized at different granularities, including model-wise separation (e.g., mixture of experts), layer-wise division (e.g., split learning), and module-wise partitioning (e.g., partitioning attention heads). Coarser partitioning strategies reduce communication cost but limit flexibility in distributing workloads, while finer-grained partitioning enables more balanced workload allocation at the cost of significant communication overhead. 
Pruning the connections within neural networks is a common technique to divide and isolate the original model into multiple components. However, this may result in a degradation of accuracy.
Designing a partitioning strategy that jointly balances computational efficiency, communication overhead, and accuracy preservation remains a central challenge for collaborative AI.

\subsubsection{Challenges from the Hardware Device}
Hardware devices in mobile edge networks encounter limitations in computation, memory, and energy capacity. From the computational perspective, support for low-precision arithmetic such as FP16, INT8, or INT4 has become essential to accelerate model inference. Achieving this capability requires joint design of hardware architectures and software drivers such as CUDA. Different precision levels inevitably lead to tradeoffs between efficiency and accuracy, which complicates deployment decisions. Memory capacity represents another critical bottleneck, as storing and executing large AI models often exceeds the resources of mobile devices. Techniques such as model partitioning and quantization can alleviate this problem. Energy constraints further restrict continuous operations on battery-powered devices. Dynamic voltage and frequency scaling (DVFS) is a commonly-adopted mechanism to balance performance and energy consumption.

\begin{table*}[t] \centering
	\caption{Spatial and temporal collaboration approaches. }\label{table1} \centering
	\begin{tabular}{clp{4.5cm}p{3.5cm}p{3.5cm}}
		\toprule
		Type & Technique & Description & Pros & Cons\\
		\midrule
		\multirow{8}{*}{Spatial}
		& Federated tuning \cite{wang2025federated}  
        & Parameter-efficient federated tuning for large models  
        &  Reduces the communication cost of parameter transmission
        &  Sensitive to data distribution across devices\\
		\cmidrule{2-5}
		& Mixture of experts \cite{wang2024toward} 
        & Activates only a subset of expert models distributed across edge devices
        & Amortizes the computational cost to multiple nodes  
        & Requires an efficient and effective routing model  \\
		\cmidrule{2-5}
		& Patch-based diffusion \cite{li2024distrifusion} 
        & Partitions input into patches for distributed parallel diffusion
        & Enables parallel diffusion inference across multiple nodes
        & Incurs additional communication between each patch\\
		\cmidrule{2-5}
		& Hierarchical diffusion \cite{yin2023nuwa} 
        & Generates global keyframes first, then refines details with local models
        & Efficient inference for long-sequence generation 
        & Incurs high synchronization cost \\
		\cmidrule{1-5}
		\multirow{8}{*}{Temporal}
		& Split learning \cite{patel2024splitwise}  
        & Splits model layers across device and server for collaborative inference
        &  Reduces the computational requirement for a single node
        &  Split point selection is critical; extra communication is needed\\
		\cmidrule{2-5}
		& Cascading learning \cite{nie2024online} 
        & Employs cascading models by deferring the hard input to larger models  
        & Achieves fast and accurate inference for easy input
        & Needs confidence calibration; deferral may increase latency \\
		\cmidrule{2-5}
		& Speculative decoding \cite{10812936} 
        & Small draft model performs generation, large model verifies and corrects them
        & Accelerates the autoregressive generation process
        &  Requires re-generation; incurs extra communication cost\\
		\cmidrule{2-5}
		& Routing inference \cite{zheng2024citer} 
        & Token-level routing between small and large models for efficient inference   
        &  Preserves quality while reducing the inference costs
        &  The routing mechanism adds system complexity\\
		\bottomrule
	\end{tabular}
\end{table*}
\subsubsection{Challenges from the Communication Networks}
Communication networks constitute another major bottleneck for collaborative AI in edge environments. Distributed inference requires frequent exchange of intermediate results between submodels or modules, which imposes non-negligible latency and bandwidth costs. Reducing this burden involves minimizing both the number of communication rounds and the amount of data transmitted per round. The former relies on the design of AI model-related mechanisms, while the latter can be achieved through compression techniques. Advanced multiple access technologies, including non-orthogonal multiple access (NOMA), provide improvements in spectrum utilization. In addition, emerging paradigms such as over-the-air computation integrate communication (AirCom) and computation by aggregating intermediate results during transmission, and semantic communication aims to transmit task-relevant information \cite{11078451}, offering new opportunities for efficiency.

\section{Resource-Efficient Collaborative Approaches}
We classify existing resource-efficient collaborative approaches into two groups: spatial collaboration and temporal collaboration. We define spatial collaboration as the parallel execution of models across multiple entities, where computation is distributed among devices simultaneously. In contrast, temporal collaboration refers to the sequential execution of models in pipeline-like structures, where the inputs of later stages depend on earlier outputs. The description, advantages, and limitations of these approaches are summarized in Table~\ref{table1}.

\subsection{Spatial Collaboration Approaches}
\subsubsection{Federated Tuning} Federated tuning enables multiple edge devices to collaboratively personalize pre-trained foundation models without sharing raw data \cite{wang2025federated}. Due to the large number of model parameters, parameter-efficient tuning methods are often integrated to reduce communication overhead. For example, low-rank adaptation introduces lightweight bypass modules, where only a small number of parameters are fine-tuned, to reduce the communication cost. Similarly, prompt tuning embeds a vector into the user prompt without updating the core parameters of the whole model.

\subsubsection{Mixture of Experts} The mixture-of-experts approach utilizes multiple expert submodels \cite{wang2024toward}, each specializing in different data regions or tasks, to mitigate single-node deployment hardware resource requirements. During inference, only a subset of experts is activated for each input, reducing computation while maintaining accuracy. In mobile edge networks, this approach allows different devices to host distinct expert models. Routing mechanisms then dynamically select the relevant experts depending on the input prompt. This distributed design reduces per-device load and enables collaborative utilization of heterogeneous resources.

\subsubsection{Patch-based Diffusion}   Diffusion models are highly effective for generative tasks but incur prohibitive costs when deployed on resource-constrained devices. Patch-based diffusion addresses this issue by spatially decomposing high-resolution inputs into local patches \cite{li2024distrifusion}, which can be distributed across multiple edge devices for parallel processing. To maintain fidelity, displaced patch parallelism can reuse pre-computed features from prior diffusion steps to current patches. This design supports asynchronous communication, which achieves significant inference acceleration while preserving global coherence in the generated outputs. 

\subsubsection{Hierarchical Diffusion} This method decomposes the generation of long videos into a hierarchical two-stage process: a global diffusion model first generates keyframes that outline the video structure, followed by parallel chunk-wise local diffusion models that fill in fine-grained details \cite{yin2023nuwa}. This design enables efficient spatial collaboration and significantly reduces the inference latency for long video generation. Nevertheless, the end-to-end latency depends on the slowest local model, which makes synchronization among different models a critical issue in practical deployments.

\subsection{Temporal Collaboration Approaches} 
\subsubsection{Split Learning} Split learning partitions a deep model across edge devices and servers, with early layers executed locally and deeper layers processed at the server \cite{patel2024splitwise}. This setup reduces device workload and memory footprint while avoiding transmission of raw data. In temporal collaboration, split learning enables devices to continuously process data streams with low latency. The position of the split point plays a critical role in overall efficiency, and it can be optimized to balance computation and communication costs. For instance, a deep split point close to the output layer may overload the client device, whereas a shallow split point may increase server workload or communication overhead.

\subsubsection{Cascading Inference}
Cascading inference employs a sequence of models with ascending complexity, where each model outputs not only a prediction but also a confidence score \cite{nie2024online}. The user input is first processed by the smallest model. When their confidence is low, queries are deferred to progressively larger models, ultimately involving LLMs if necessary. The cascade learning framework can reduce the computation cost while maintaining accuracy comparable to full LLM inference. Simple queries benefit from rapid responses, whereas complex queries may undergo multiple deferrals, leading to increased inference latency.

\subsubsection{Speculative Decoding} Autoregressive generation with large models is slow due to sequential token prediction. Speculative decoding accelerates this process by employing a smaller draft model to generate candidate tokens, which are then verified by the larger target model, with a probability score assigned to each token \cite{10812936}. The verification step is computationally efficient and can be conducted in parallel. If the target model assigns low probability to any draft tokens (i.e., mismatches occur), it resamples and corrects only those disputed tokens. To further improve performance, speculative decoding is typically combined with adaptive fallback mechanisms and pipelined verification strategies. 

\subsubsection{Routing Inference}
Routing inference leverages token-level collaboration between small and large models \cite{zheng2024citer}. Instead of applying the same model uniformly across the entire input sequence, a routing policy determines which tokens are simple and can be handled by a lightweight model, and which tokens are more complex and should be processed by a larger model. Training such a routing policy often relies on reinforcement learning, where the objective is to optimize both prediction accuracy and resource efficiency. This approach preserves high generation quality while reducing inference overhead, offering a promising solution for real-time mobile applications.

\subsection{Lessons Learned}
By analyzing both spatial and temporal collaboration, we identify several lessons that can guide the design of resource-efficient collaborative AI systems as follows.

\subsubsection{No Universal Paradigm} 
There is no single collaboration paradigm that can apply universally across all scenarios. The optimal approach depends on task characteristics, input/output structure, and available resources. Spatial collaboration excels in parallelizable, computation-intensive workloads with minimal inter-task dependencies, where efficiency is influenced by the granularity of submodel partitioning. Temporal collaboration is preferable for tasks with strong sequential dependencies, and its end-to-end latency is determined by the slowest stage in the pipeline.

\subsubsection{Communication Bottleneck} 
Communication overhead remains the key bottleneck in collaborative AI. In spatial collaboration, the efficiency of parallel deployment is bounded by the slowest submodel, thereby introducing synchronization issues. In temporal collaboration, token- or feature-level exchanges between small and large models are lightweight, but repeated interactions can accumulate into substantial delays. Therefore, efficient communication and synchronization protocols are critical to improving overall performance \cite{11078451}.

\subsubsection{Hybrid Designs}
Spatial and temporal collaboration are not mutually exclusive, and hybrid strategies can exploit their complementary advantages. For instance, a mixture-of-experts framework can be integrated with speculative decoding. Specifically, experts perform specialized subtasks in parallel, while speculative decoding accelerates token-level generation within each expert’s inference. Such hybrid approaches offer a promising pathway toward achieving both scalability and responsiveness in cloud-edge-client collaborative AI deployments.

\begin{figure*}[t]\centering
  \includegraphics[width=0.76\textwidth]{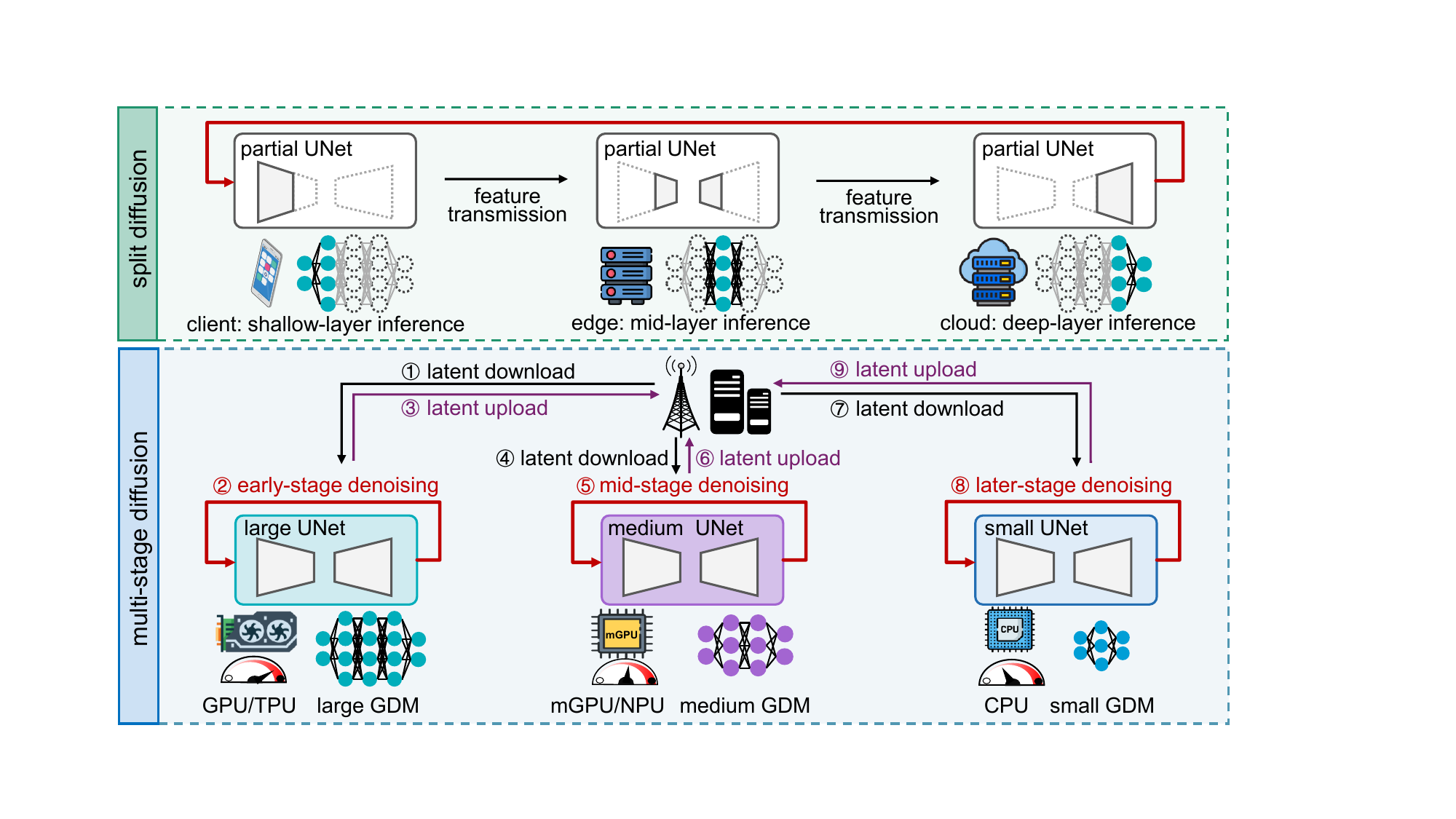}
  \caption{Illustrations of split diffusion and edge-assisted multi-stage diffusion. \textbf{Top}: Split diffusion partitions the model across devices, requiring repeated transmission of feature data at each denoising step. 
  \textbf{Bottom}: Multi-stage diffusion divides the process into contiguous stages, exchanging only latent data between stages to reduce the communication cost.}\label{fig:relayf}
  \vspace{-9pt}
\end{figure*}

\section{Case Study of Collaborative Diffusion}
\subsection{Edge-Assisted Multi-Stage Diffusion}
The efficient deployment of GDMs is critical for mobile edge networks \cite{10454003}. On the one hand, GDMs can be applied to generating high-quality images and videos for immersive experiences. On the other hand, GDMs can enable intelligent decision-making when integrated with reinforcement learning frameworks.
To accelerate the diffusion process in resource-constrained environments, a common distributed strategy is split learning, which partitions the model across multiple devices, as illustrated at the top of Figure \ref{fig:relayf}. However, this strategy is poorly suited to the iterative nature of diffusion processes, as each iteration requires full-model execution, leading to repeated transmission of high-dimensional intermediate feature maps. Consequently, this results in excessive communication overhead and increased latency, which undermines the benefits of distributed computation.

To overcome the above limitations, we propose an edge-assisted multi-stage diffusion framework, which is illustrated in Figure~\ref{fig:relayf}. The framework comprises two core components, i.e., a distributed diffusion strategy and a device selection scheme. Instead of splitting the model vertically, we divide the diffusion process horizontally into multiple stages, each consisting of several denoising steps. Each stage is assigned to one edge device, which performs a contiguous block of inference steps. The edge server relays the latent representations between devices, reducing communication frequency compared to per-step transmission.

We first design a distributed diffusion strategy that partitions the diffusion process horizontally across devices.
Formally, consider a set of devices $\mathcal{I}=\{1,2,\ldots,I\}$, coordinated by the edge server. Each device maintains a local diffusion model of size $S_i$, and we aim to select a subset $\mathcal{J}\subseteq\mathcal{I}$ for collaboration. For each selected device $j\in\mathcal{J}$, the following operations are performed in order.
\begin{itemize}
    \item The device downloads a latent representation from the edge server, incurring latency of $T_{j}^{\text{down}}$ and energy of $E_{j}^{\text{down}}$, both dependent on channel conditions and data size. When $j=1$, the latent corresponds to initial noise, serving as the diffusion starting point.
    \item The device then performs $K$ denoising steps with its local model, consuming energy of $E_{j}^{\text{cmp}}$  and latency of $T_{j}^{\text{cmp}}$. This constitutes the primary workload in the pipeline.
    \item Afterward, the refined latent is uploaded to the edge server, incurring latency of $T_{j}^{\text{up}}$ and energy of $E_{j}^{\text{up}}$. The edge server then relays the refined latent to the next device $j+1$ in $\mathcal{J}$ to continue the inference.
\end{itemize}

Specifically, the edge server maintains the most recent latent representations as the checkpoint for all selected devices. If a device becomes unavailable, the framework can restore the latest valid checkpoint and reassign the remaining denoising stages to alternative devices, ensuring uninterrupted completion of the diffusion process.
We then propose a device selection scheme that determines the optimal subset $\mathcal{J}$ by solving a constrained optimization problem, aiming to maximize generative quality subject to the system-wise latency and energy constraints. The explanations are detailed as follows.
\begin{itemize}
\item Generative quality is influenced by the capacity of the local models. Since model performance is positively correlated with model size, we directly maximize the total model size, expressed as $\sum_{j\in \mathcal{J}} S_j$.
\item  The total latency is obtained by summing the downloading, computation, and uploading times of all $J$ selected devices. The end-to-end latency is constrained by a predefined bound $T_{\max}$.
\item The overall energy cost is the sum of the energy expenditures of the selected devices, including communication and computation. This total consumption has to remain below a predetermined energy budget $E_{\max}$.
\end{itemize}

It can be verified that the above optimization problem can be viewed as a multi-dimensional knapsack problem, whose computational complexity is NP-hard. In practice, the number of available edge devices $I$ is typically moderate, and the ranges of $T_{\max}$ and $E_{\max}$ are relatively small. Therefore, we can solve this problem via a constraint-discretized dynamic programming (DP) algorithm to obtain a near-optimal solution within acceptable complexity.

\begin{itemize}
\item We discretize the latency and energy budgets into integer capacities, and construct a three-dimensional DP table.
\item We update the DP table in a nested manner over devices and resource budgets. For each device $j$, we decide whether to skip or select it. 
\item After filling the DP table, we obtain the optimal total model size and recover the corresponding device subset $\mathcal{J}$ via backtracking.
\end{itemize}

Beyond the latency–energy–quality optimization, fairness among participating devices is a crucial issue. For instance, repeatedly selecting only high-performance devices may lead to disproportionate energy depletion or long-term workload imbalance. Future work can also incorporate fairness-aware mechanisms to ensure equitable resource utilization.

\begin{figure*}[t]\centering
  \includegraphics[width=0.82\textwidth]{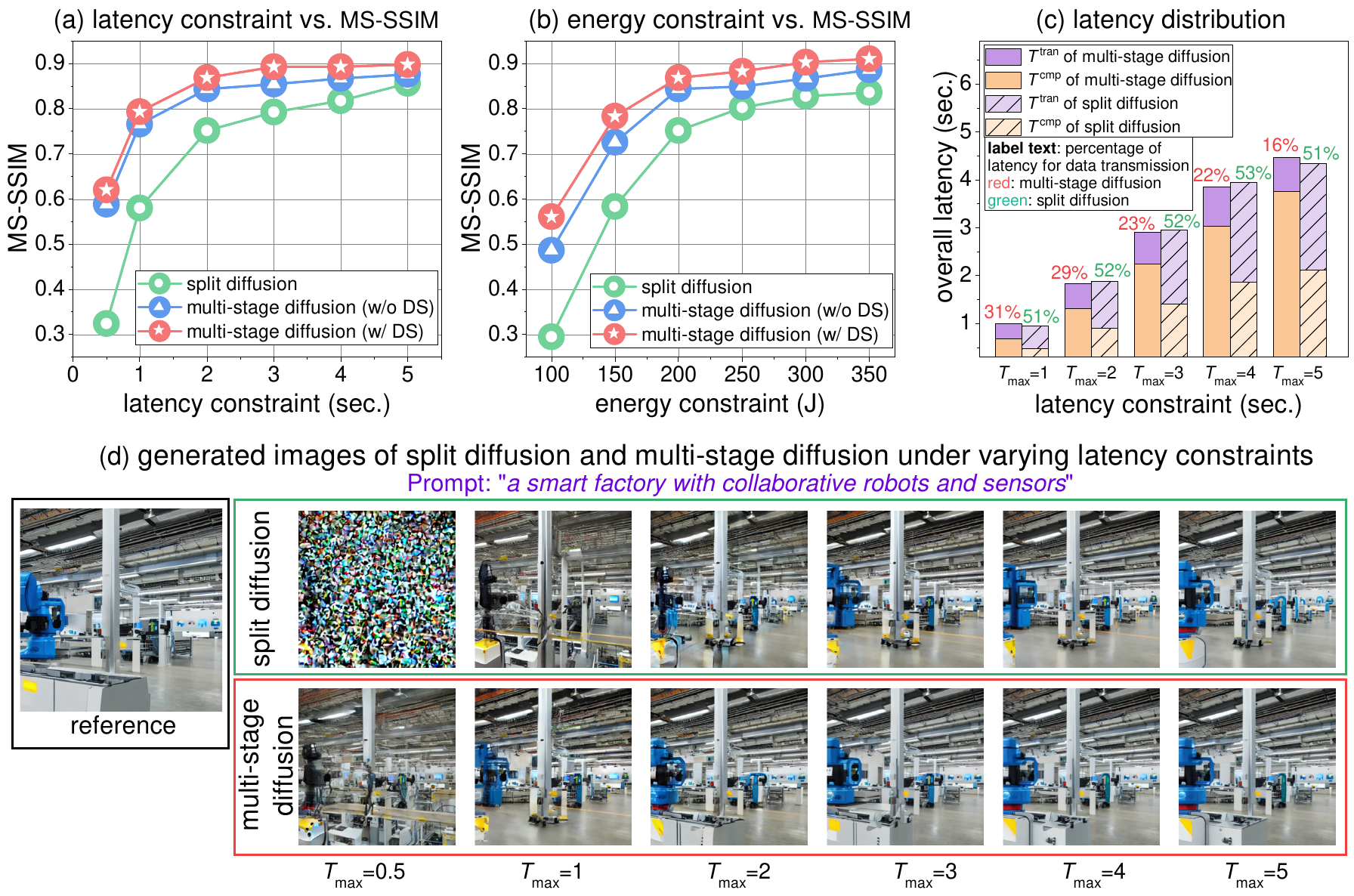}
  \caption{Experiments comparison of split diffusion and multi-stage diffusion. (a-b): quality of the generated images under varying resource constraints; (c): proportions of latency incurred by data transmission and computation, respectively; (d) visualization of the generated images by different methods.}\label{fig:exp}
  \vspace{-8pt}
\end{figure*}

\subsection{Performance Evaluation}
To evaluate the effectiveness of the proposed framework, we conduct simulations with $I=20$ edge devices, including Jetson AGX Xavier and Jetson Orin devices. These devices are randomly distributed within a $500 \times 500$ m$^2$ cellular area and coordinated by an edge server equipped with an NVIDIA A5000 GPU. 
We adopt Stable Diffusion v1.5\footnote{\url{https://huggingface.co/stable-diffusion-v1-5/stable-diffusion-v1-5}} as the base generative diffusion model. We apply different quantization strategies to obtain four model variants, including 32-bit, 16-bit, 8-bit, and 4-bit, and then randomly assign these models to local devices. 
For the computation, each selected device will perform a 5-step diffusion inference.
For the data transmission, the latent representation exchanged between two consecutive devices has a size of 256 Kbytes.
For the purpose of comparison, we implement a two-segment split inference scheme, in which the diffusion model is partitioned into two submodels executed on separate devices. The maximum energy consumption and maximum latency constraints are respectively set as $E_{\max}=200$ J and $T_{\max}=2$ seconds by default. 

Figure \ref{fig:exp}(a) and Figure \ref{fig:exp}(b) present the quality of the generated images, measured using the mean structural similarity index (MS-SSIM). A higher MS-SSIM value corresponds to better image quality. As the available resources increase (i.e., $T_{\max}$ and $E_{\max}$ increase), the quality of the generated images improves. More importantly, under the same $T_{\max}$ and $E_{\max}$, the proposed multi-stage diffusion method consistently outperforms the split diffusion approach. Furthermore, the management strategy incorporating device selection (w/ DS) yields superior performance compared to the strategy without device selection (w/o DS). This improvement is attributed to the optimized resource allocation across heterogeneous devices by leveraging the diversity of edge devices.

Figure \ref{fig:exp}(c) further investigates the time distribution between data transmission and computation for both multi-stage diffusion and split diffusion under varying latency constraints. As $T_{\max}$ increases, the proportion of time allocated to data transmission $T^{\text{tran}}$ decreases for multi-stage diffusion. This is attributed to the efficient communication design of the multi-stage diffusion framework, which decouples transmission overhead from denoising steps. In contrast, for split diffusion, the communication overhead remains a major bottleneck even as the latency constraint is relaxed, resulting in a high proportion of time spent on data transmission. This demonstrates the advantage of the multi-stage framework, which reduces communication costs while maintaining high generative quality.

Figure \ref{fig:exp}(d) shows the images generated by both multi-stage diffusion and split diffusion under different time constraints. All images are generated from the same prompt and initialized with identical latent noise. The multi-stage diffusion approach maintains high fidelity in the output. In contrast, the images generated by split diffusion exhibit more artifacts and distortion, particularly under stricter latency constraints. This highlights the robustness of the multi-stage diffusion framework in generating high-quality images, even under
varying system conditions.

\section{Open Research Directions}
There are several open directions to further enhance the efficiency and applicability of the collaborative AI system. The details are illustrated as follows.

1) \textit{Large Model Assisted Semantic Communication:} The advent of large models opens up new possibilities for semantic communication, which surpass traditional methods in encoding and decoding capabilities. A potential approach is to leverage smaller models at client devices to handle encoding tasks, while large models at edge/cloud servers perform the decoding. This asymmetrical-model-based semantic transmission system can ensure efficient resource utilization of heterogeneous resources while maintaining high-fidelity data reconstruction at the receiver side. Furthermore, token-level communication using LLMs or iterative decoding with GDMs could provide additional gains in efficiency and accuracy.

2) \textit{Privacy Protection for Large Model Collaboration:} The collaborative nature of AI systems often requires the exchange of intermediate data (e.g., latents and tokens), raising significant privacy concerns. One promising solution is the integration of differential privacy techniques to secure data during transmission. Additionally, model-specific security techniques are essential to protect against threats like prompt injections and jailbreak chains targeting large models. To ensure comprehensive privacy protection, the collaborative defense framework should be developed, where small models on the input side filter malicious input prompts, while large models on the output side mitigate harmful content. 

3) \textit{Large Model Collaboration for Low Altitude Networks:} Low-altitude networks, incorporating uncrewed aerial vehicles (UAVs), generate rich multimodal data that can be efficiently processed using collaborative large AI model frameworks. In these networks, small models deployed on UAVs handle real-time local processing tasks, such as object detection and terrain mapping, while larger models at ground stations manage complex decision-making and global analysis. This collaboration enhances the intelligence and autonomy of low-altitude networks, particularly in applications like autonomous navigation and airspace management. 

4) \textit{Sustainable Collaborative AI System:} Energy consumption has become a crucial bottleneck in collaborative AI systems, where mobile devices rely on battery power for computational tasks. Collaborative AI systems may necessitate frequent information exchange between edge servers and mobile devices. Simultaneous wireless information and power transfer (SWIPT) addresses the energy challenge by enabling edge servers to transmit both data and power simultaneously.
For example, when an edge server sends intermediate data to a mobile device, SWIPT can simultaneously charge the device, ensuring that it has sufficient power to perform local model inference tasks.

\section{Conclusion}
In this article, we have proposed the framework of collaborative AI in mobile edge networks, which enables the distribution of large AI models across heterogeneous edge devices to enhance inference efficiency and service quality. We have reviewed several key resource-efficient collaboration approaches and highlighted their respective advantages. We have then discussed the major challenges in implementing collaborative AI in mobile edge networks. We have proposed a multi-stage diffusion framework to optimize the collaboration of generative models. Finally, we have illustrated several open research directions and open problems, focusing on improving the efficiency and scalability of collaborative AI systems in mobile edge networks.

\bibliographystyle{IEEEtran}
\bibliography{reference.bib}

\end{document}